\documentstyle[12pt]{article}
\begin{document}
{\pagestyle{empty}
\rightline{TU-02/96}
\rightline{Feb. 1996}
\rightline{~~~~~~~~~}
\vskip 1cm
\centerline{\Large \bf $N=2$ and $4$ Super Yang-Mills Theories}
\centerline{\Large \bf on $M_4 \times Z_2 \times Z_2$ Geometry}
\vskip 1cm

\centerline{{Bin Chen\footnote{E-mail address: cb@itp.ac.cn}},
            {Takesi Saito\footnote{E-mail address: 
                         tsaito@jpnyitp.yukawa.kyoto-u.ac.jp}},
            {Hong-Bo Teng\footnote{E-mail address: tenghb@itp.ac.cn}},}
\centerline{{Kunihiko Uehara\footnote{E-mail address:                                      uehara@tezukayama-u.ac.jp}}
        and {Ke Wu\footnote{E-mail address: wuke@itp.ac.cn}}}
\vskip 1cm
\centerline{\it Institute of Theoretical Physics, Academia Sinica, P.O.Box 2735, Beijing 100080, China${}^{1,3,5}$}
\centerline{\it Department of Physics, Kwansei Gakuin University, Nishinomiya 662, Japan${}^2$}
\centerline{\it Department of Physics, Tezukayama University, Nara 631, Japan${}^4$}

\vskip 2cm

\centerline{\bf Abstract}
\vskip 0.2in

We derive the $N=2$ and $4$ super Yang-Mills theories from the viewpoint of 
the $M_4\times Z_2\times Z_2$ gauge theory. Scalars and pseudoscalars 
appearing in the theories are regarded as gauge fields along the directions 
on $Z_2\times Z_2$ discrete space.

\vskip 0.4cm
\noindent
PACS number(s): 
\hfil
\vfill
\newpage}
\renewcommand{\theequation}{\thesection.\arabic{equation}}
\setcounter{equation}{0}
\addtocounter{section}{1}
\section*{\S\thesection.\ \ Introduction}

\indent

  The non-commutative geometric construction of Connes\cite{Connes1,Connes2} 
has been successful in giving a geometrical interpretation of the standard 
model as well as some grand unification models.  
In this interpretation the Higgs fields are regarded as fields along 
directions in the discrete space.  
The bosonic parts of the actions are just the pure Yang-Mills actions 
containing gauge fields on both continuous and discrete spaces, and the 
Yukawa coupling is regarded as a kind of gauge interactions of fermions.

  At the same time, applying non-commutative geometry(NCG) to SUSY theories 
has encountered many difficulties.  
A natural way is to introduce a non-commutative space which is a product of 
the superspace and a set of discrete points, similar to those which have been 
done in non-SUSY theories.  
However, such an extension of superspace has been proved to be rather 
difficult to accomplish.  
Chamseddine\cite{Chamseddine} then proposed an alternative approach in which 
SUSY theories were considered in their component form.  
He discussed how to derive from NCG the $N=2$ and $4$ SUSY Yang-Mills actions, 
and also the coupling of $N=1$ and $2$ super Yang-Mills fields to 
$N=1$ and~$2$ matters.

  Chamseddine's paper is the first one in which a connection between 
space-time supersymmetry and NCG is discussed.  
However his approach is rather complicated, especially the geometric meaning 
of $Z_2$ he used is not so clear.  
In our paper we use $Z_2 \times Z_2$ rather than $Z_2$ only.  
Then we would like to discuss how to derive the $N=2$ and $4$ SUSY Yang-Mills 
theories from the viewpoint of the $M_4 \times Z_2 \times Z_2$ gauge theory,
which was previously proposed by Konisi and Saito\cite{Konisi} without 
recourse to NCG.
This approach appears to be geometrically very simple and clear.
The scalar fields $S^a(x)$, $P^a(x)$ in $N=2$ theory and $A^{aI}(x)$, 
$B^{aI}(x)$ ($I=1,2,3$) in $N=4$ theory will be regarded as gauge fields along 
directions on $Z_2 \times Z_2$ discrete space.  

  This paper is scheduled as follows: \ \ In \S 2 we summarize the extended 
gauge theory on $M_4 \times Z_2 \times Z_2$ without recourse to 
NCG\cite{Konisi}.  
This will be applied to $N=2$ theory in \S 3 and to 
$N=4$ theory in \S 4, respectively.  
The final section is devoted to concluding remarks.  

\setcounter{equation}{0}
\addtocounter{section}{1}
\section*{\S\thesection.\ \ Gauge theory on $M_4 \times Z_2 \times Z_2$}

\indent

  In this section we summarize the gauge theory on $M_4 \times Z_2 \times Z_2$ \cite{Konisi}.
  Let us write the four elements of $Z_2 \times Z_2$ as 
\begin{equation}
g_0=(e_1,e_2), \ g_1=(r_1,e_2), \ g_2=(e_1,r_2), \ g_3=(r_1,r_2). \label{e201}
\end{equation}
\noindent
They are subject to the algebra
\begin{eqnarray}
g_0+g_i=g_i,&& \ g_i+g_i=g_0,\ (i=1,2,3)\nonumber \\
g_1+g_2=g_3 && {\rm\ and\ cyclic}.
\label{e202}
\end{eqnarray}

To every point $(x,p)$ with $x\in M_4$ and $p=g_0, g_1, g_2, g_3$ we attach 
a complex $n$-dimensional internal vector space $V_n[x,p]$.  Generally, $n$ may 
take different values with each other for different $p$'s.  However, we 
confine ourselves here in the equal $n$-dimensional case.  For any scalar 
field $f(x,p)$ on $V_n[x,p]$ we define the difference $\delta_hf(x,p)$ by 
\begin{eqnarray}
\delta_hf(x,p)&=&f(x,p)-f(x,p+h),\nonumber \\
             h&=&g_1,g_2,g_3.
\label{e203}
\end{eqnarray}

\noindent
It is easy to check the identity 
\begin{equation}
\delta_k \delta_h f(x,p)-C_{kh}^{\,l} \delta_l f(x,p)=0, \ 
C_{kh}^{\,l}=\delta_k^l+\delta_h^l-\delta_{k+h}^{\,l}.
\label{e204} 
\end{equation}

\noindent
Namely, the second-order difference can be written by the first-order 
differences.

  For the fermion field $\psi^a(x,p)$ which is a vector on $V_n[x,p]$, 
Eq.(\ref{e203}) should be replaced by a covariant difference 
defined by
\begin{equation}
\nabla\!_h\psi^a(x,p)=\psi^a(x,p)-(H(x,p,p+h))^a_{\,b}\psi^b(x,p+h), 
\label{e205}
\end{equation}

\noindent
where $a,b=1,2,\cdots,n$.  Since $\psi^a(x,p)$ and $\psi^b(x,p+h)$ are 
different vectors belonging to different internal spaces with each other, 
so the simple difference $\psi^a(x,p)-\psi^a(x,p+h)$ has no meaning.  However, 
if we give a scalar function $H(x,p,p+h)^a_{\,b}$ of $n \times n$ matrix, the 
$\psi^a(x,p+h)$ is mapped to vector $\psi^a_H(x,p)$ on $V_n[x,p]$ by 
the product $\psi_H^a(x,p)=H(x,p,p+h)^a_{\,b}\psi^{\,b}(x,p+h)$, where $H(x,p,p+h)$ is 
subject to a rule of the gauge transformation 
\begin{equation}
H(x,p,p+h) \rightarrow \widetilde H(x,p,p+h)=U^{-1}(x,p)H(x,p,p+h)U(x,p+h)
\label{e206}
\end{equation}

\noindent
under a rotation $U(x,p)$ of the $V_n[x,p]$ frame
\begin{equation}
\psi^a(x,p) \rightarrow 
\widetilde \psi^a(x,p)=(U^{-1}(x,p))^a_{\,a'}\psi^{a'}(x,p).
\label{e207}
\end{equation}

\noindent
Generally such a mapping function $H(x,p,p+h)$ is called a connection, and 
it is, therefore, regarded as the gauge field associated with $Z_2 \times Z_2$.
Henceforth we refer to 
$\psi^a_H(x,p)=H(x,p,p+h)^a_{\,b}\psi^b(x,p+h)$ as 
the parallel-transported vector of $\psi^b(x,p+h)$ from $p+h$ to $p$.

  In order to define a field strength (or curvature) for $H(x,p,p+h)$ 
we calculate the commutator
\begin{equation}
[\nabla\!_k,\nabla\!_h]\psi^a(x,p)=-(\widetilde F_{kh}(x,p))^a_{\,b}\psi^b(x,p+k+h),
\ (h,k=g_1,g_2,g_3)
\label{e208}
\end{equation}

\noindent
where
\begin{equation}
\widetilde F_{kh}(x,p)=-[H(x,p,p+k)H(x,p+k,p+k+h)-H(x,p,p+h)H(x,p+h,p+k+h)].
\label{e209}
\end{equation}

\noindent
This function $\widetilde F_{kh}(x,p)$ can be regarded as such a field 
strength.  The reason is as follows:  The first term 
$H(x,p,p+k)H(x,p+k,p+k+h)\psi(x,p+k+h)$ shows the parallel-transported 
vector of $\psi(x,p+k+h)$ from $p+k+h$ to $p$ through $p+k$ (see Fig.A), 
whereas the second term $H(x,p,p+h)H(x,p+h,p+k+h)\psi(x,p+k+h)$ shows 
the parallel-transported vector of $\psi(x,p+k+h)$ from $p+k+h$ to $p$ 
through another point $p+h$.  The difference between both parallel 
transportations will, therefore, give the curvature 

  However, we can consider another type of parallel transportations depicted 
in Fig.B, because on the discrete space any second-order difference can be 
written by the first-order one as was shown in (\ref{e204}).  Actually, such 
a difference of the parallel transportations is given by 
\begin{equation}
(\nabla\!_k\nabla\!_h-C_{kh}^{\,l}\nabla\!_l)\psi^a(x,p)=
-(F_{kh}(x,p))^a_{\,b}\psi^b(x,p+k+h),
\label{e210}
\end{equation}

\noindent
where
\begin{equation}
F_{kh}(x,p)=H(x,p,p+k+h)-H(x,p,p+k)H(x,p+k,p+k+h).
\label{e211}
\end{equation}

\noindent
Namely, the first term $H(x,p,p+k+h)\psi(x,p+k+h)$ shows the parallel-%
transported vector of $\psi(x,p+k+h)$ from $p+k+h$ to $p$ directly, whereas 
the second term $H(x,p,p+k)H(x,p+k,p+k+h)\psi(x,p+k+h)$ shows 
the parallel-transported vector of $\psi(x,p+k+h)$ from $p+k+h$ to $p$ 
through $p+k$.  The difference between such both parallel transportations 
will give another curvature $F_{kh}(x,p)$.  Henceforth we call it the triangle 
curvature.

\setlength{\unitlength}{1mm}
\begin{picture}(165,40)(0,0)
 \put(50,10){\vector(-1,1){6}}\put(44,16){\line(-1,1){4}}
 \put(40,20){\vector(1,1){6}}\put(46,26){\line(1,1){4}}
 \put(50,10){\vector(1,1){6}}\put(56,16){\line(1,1){4}}
 \put(60,20){\vector(-1,1){6}}\put(54,26){\line(-1,1){4}}
 \put(50,10){\circle*{1}}
 \put(40,20){\circle*{1}}
 \put(60,20){\circle*{1}}
 \put(50,30){\circle*{1}}
 \put(49,32){$p$}
 \put(29,19){$p+h$}
 \put(62,19){$p+k$}
 \put(42,06){$p+k+h$}
 \put(25,06){{\bf Fig.A.}}
 \put(100,10){\vector(0,1){11}}\put(100,21){\line(0,1){9}}
 \put(100,10){\vector(1,1){6}}\put(106,16){\line(1,1){4}}
 \put(110,20){\vector(-1,1){6}}\put(104,26){\line(-1,1){4}}
 \put(100,10){\circle*{1}}
 \put(110,20){\circle*{1}}
 \put(100,30){\circle*{1}}
 \put(99,32){$p$}
 \put(112,19){$p+k$}
 \put(99,06){$p+k+h$}
 \put(82,06){{\bf Fig.B.}}
\end{picture}

\noindent
Two kinds of curvature $\widetilde F_{kh}(x,p)$ and $F_{kh}(x,p)$ have a 
relation
\begin{equation}
\widetilde F_{kh}(x,p)=F_{kh}(x,p)-F_{hk}(x,p),
\label{e212}
\end{equation}

\noindent
namely, $\widetilde F_{kh}(x,p)$ corresponds to an antisymmetric part of 
$F_{kh}(x,p)$.

  The ordinary Yang-Mills field $\omega_\mu(x,p)$ is introduced by the covariant 
derivative
\begin{equation}
\nabla\!_\mu\psi^a(x,p)=(\partial_\mu+i\,\omega_\mu(x,p))^a_{\,b}\psi^b(x,p). \ (\mu=0,1,2,3)
\label{e213}
\end{equation}

\noindent
We assume that $\omega_\mu(x,p)$ is independent of $p$ and is set to be
\begin{equation}
\omega_\mu(x,p)=A_\mu(x).
\label{e214}
\end{equation}

\noindent
Its curvature is given by
\begin{equation}
[\nabla\!_\mu,\nabla\!_\nu]\psi^a(x,p)=
		i\,(F_{\mu\nu}(x))^a_{\,b}\psi^b(x,p),
\label{e215}
\end{equation}

\noindent
where
\begin{equation}
F_{\mu\nu}(x)=\partial_\mu A_\nu(x)-\partial_\nu A_\mu(x)
	+i\,[A_\mu(x),A_\nu(x)].
\label{e216}
\end{equation}

The other curvature component $F_{\mu h}(x,p)$ is calculated to be
\begin{equation}
[\nabla\!_\mu,\nabla\!_h]\psi^a(x,p)=-F_{\mu h}(x,p)^a_{\,b}\psi^b(x,p+h),
\label{e217}
\end{equation}

\noindent
where
\begin{equation}
F_{\mu h}(x,p)=\partial_\mu H(x,p,p+h)
	+i\,[A_\mu(x),H(x,p,p+h)]=\nabla\!_\mu H(x,p,p+h).
\label{e218}
\end{equation}

\noindent
Here, we need no accounting for a triangle-like curvature, since $[\partial_\mu,\delta_h]=0$ for any function.

  By taking into account of four kinds of curvatures the bosonic Lagrangian is now given by
\begin{equation}
{\cal L}_{\rm B}={\cal L}_1+{\cal L}_2+{\cal L}_3
\label{e219}
\end{equation}

\noindent
with
{\addtocounter{equation}{-1}
\setcounter{enumi}{\value{equation}}
\addtocounter{enumi}{1}
\setcounter{equation}{0}
\renewcommand{\theequation}{\thesection.\theenumi\alph{equation}}
\begin{eqnarray}
{\cal L}_1&=&-\frac{1}{4}F^a_{\mu\nu}(x)F^{\mu\nu}_a(x),
\label{e219a} \\
{\cal L}_2&=&\xi\,\displaystyle{\sum_p}{\rm tr}[F_{\mu h}^{\dag}(x,p)
              F^{\mu h}(x,p)] \nonumber \\
          &=&\xi\,\displaystyle{\sum_{p,h}}{\rm tr}
              [(\nabla\!_\mu H(x,p,p+h))^{\dag}(\nabla^\mu H(x,p,p+h))],
\label{e219b} \\
{\cal L}_3&=&\eta\,\displaystyle{\sum_p}{\rm tr}[F_{kh}^{(S)\dag}(x,p)
              F^{(S)kh}(x,p)]
            +\zeta\,\displaystyle{\sum_p}{\rm tr}[F_{kh}^{(A)\dag}(x,p)
              F^{(A)kh}(x,p)],
\label{e219c}
\end{eqnarray}
\setcounter{equation}{\value{enumi}}
}

\noindent
where $\xi$, $\eta$ and $\zeta$ are real normalization constants, 
$F_{kh}^{(S)}$ and $F_{kh}^{(A)}$ are symmetric and antisymmetric parts of 
the triangle curvature $F_{kh}$, respectively.

  A fermionic Lagrangian may be written as
\begin{equation}
{\cal L}_{\rm F}=i\,\displaystyle{\sum_p}\bar\psi_a(x,p)(\Gamma^\mu \nabla\!_\mu+\Gamma^h \nabla\!_h)\psi^a(x,p), 
\label{e220}
\end{equation}

\noindent
where
\begin{equation}
\Gamma^\mu=\gamma^\mu \times \frac{\tau^0}{2},\ 
\Gamma^h=\gamma_5 \times \frac{\tau^h}{2},\ 
(h=g_1, g_2, g_3 {\rm\ or\ simply\ } 1,2,3)
\label{e221}
\end{equation}

\noindent
$\tau^0$ being a $(2\times 2)$ unit matrix and $\tau^h$ the Pauli matrix.

\setcounter{equation}{0}
\addtocounter{section}{1}
\section*{\S\thesection.\ \ $N=2$ super Yang-Mills theory}

\indent

  The $N=2$ super Yang-Mills action\cite{Salam} is known to be
\begin{eqnarray}
I_2&=\displaystyle{\int}d^4x\Bigl[&\!-\frac{1}{4}F_{\mu\nu}^aF^{\mu\nu}_a
     +\frac{1}{2}\nabla\!_\mu S^a\nabla^\mu S_a
     +\frac{1}{2}\nabla\!_\mu P^a\nabla^\mu P_a
     +i\bar\chi^a\gamma^\mu \nabla\!_\mu \chi_a \nonumber\\
   & &-if_{abc}\bar\chi^a(S^b+i\gamma_5 P^b)\chi^c
     -\frac{1}{2}(f_{abc}S^b P^c)^2\Bigr],
\label{e301}
\end{eqnarray}

\noindent
where $S^a$ and $P^a$ are scalar and pseudoscalar fields, respectively, 
and $\chi^a$ is a Dirac spinor, all in the adjoint representation of the 
gauge group $G$ with the structure constant $f_{abc}$.  The action is 
invariant under the $N=2$ super transformations.  Our purpose is to consider 
a relationship between the above theory and the $M_4\times Z_2\times Z_2$ 
gauge theory.

  In the fermionic Lagrangian (\ref{e220}) we require
\begin{equation}
\nabla\!_3\psi^a(x,p)=\psi^a(x,p)-(H(x,p,p+g_3))^a_{\,b}\psi^b(x,p+g_3)=0
\label{e302}
\end{equation}

\noindent
and 
\begin{equation}
(H(x,p,p+g_3))^a_{\,b}=\delta^a_{\,b}.
\label{e303}
\end{equation}

\noindent
From these it follows that
\begin{equation}
\psi(x,p)=\psi(x,p+g_3),
\label{e304}
\end{equation}

\noindent
hence
\begin{equation}
\psi(x,g_0)=\psi(x,g_3){\rm\ \ and\ \ }\psi(x,g_1)=\psi(x,g_2).
\label{e305}
\end{equation}

\noindent
From (\ref{e303}) we have
\begin{equation}
H(x,p,p+g_3)=H(x,p+g_1,p+g_2)=H(x,p+g_2,p+g_1)=H(x,p+g_3,p)=1
\label{e306}
\end{equation}

\noindent
For other covariant differences
\begin{eqnarray}
\nabla\!_1\psi(x,p)=\psi(x,p)-H(x,p,p+g_1)\psi(x,p+g_1),
\label{e307} \\
\nabla\!_2\psi(x,p)=\psi(x,p)-H(x,p,p+g_2)\psi(x,p+g_2),
\label{e308}
\end{eqnarray}

\noindent
we set
\begin{eqnarray}
H(x,p,p+g_1)=P(x)=T_aP^a(x),
\label{e309} \\
H(x,p,p+g_2)=S(x)=T_aS^a(x),
\label{e310}
\end{eqnarray}

\noindent
where $T^a$ is the generator of $G$ subject to algebra
\begin{equation}
[T_a,T_b]=if_{abc}T^c.
\label{e311}
\end{equation}

\noindent
By substituting $p+g_i$ into $p$ in (\ref{e309}) and (\ref{e310}) we find 
\begin{eqnarray}
H(p,p+g_1)=H(p+g_1,p)=H(p+g_2,p+g_3)=H(p+g_3,p+g_2)=P(x),
\label{e312} \\
H(p,p+g_2)=H(p+g_1,p+g_3)=H(p+g_2,p)=H(p+g_3,p+g_1)=S(x).
\label{e313}
\end{eqnarray}

\noindent
If we put in (\ref{e305})
\begin{equation}
\psi^{a}(x,g_0)=\psi^{a}(x,g_3)={L\chi^a\choose0}
{\rm\ \ and\ \ }
\psi^{a}(x,g_1)=\psi^{a}(x,g_2)={0\choose R\chi^a},
\label{e314}
\end{equation}

\noindent
where $L$ and $R$ are left-handed and right-handed projection operators, 
respectively, {\it i.e.},
\begin{equation}
L=\frac{1-\gamma_5}{2}{\rm\ \ and\ \ }R=\frac{1+\gamma_5}{2},
\label{e315}
\end{equation}

\noindent
then the fermionic Lagrangian (\ref{e220}) is reduced to 
\begin{equation}
{\cal L}_{\rm F}=i\,\bar\chi^a(x)\gamma^\mu\nabla\!_\mu\chi_a(x)
          -\bar\chi^a(x)[S(x)+i\gamma_5 P(x)]_{ab}\chi^b(x). 
\label{e316}
\end{equation}

\noindent
This is equivalent to that in (\ref{e301}) in the adjoint representation 
$((T_a)_{bc}=-if_{abc})$.

  Next we consider the bosonic Lagrangian.  
The triangle curvature (\ref{e211}) is given by
\begin{equation}
F_{ij}(p)=H(p,p+g_i+g_j)-H(p,p+g_i)H(p+g_i,p+g_i+g_j),
\label{e317}
\end{equation}

\noindent
so that
{\addtocounter{equation}{-1}
\setcounter{enumi}{\value{equation}}
\addtocounter{enumi}{1}
\setcounter{equation}{0}
\renewcommand{\theequation}{\thesection.\theenumi\alph{equation}}
\begin{eqnarray}
F_{12}(p)&=&H(p,p+g_3)-H(p,p+g_1)H(p+g_1,p+g_3)=1-P(x)S(x),
\label{e317a} \\
F_{21}(p)&=&H(p,p+g_3)-H(p,p+g_2)H(p+g_2,p+g_3)=1-S(x)P(x),
\label{e317b} \\
F_{23}(p)&=&F_{32}(p)=-F_{31}(p)=-F_{13}(p)=-S(x)+P(x),
\label{e317c} \\
F_{11}(p)&=&1-P^2(x),
\label{e317d} \\
F_{22}(p)&=&1-S^2(x),
\label{e317e} \\
F_{33}(p)&=&0.
\label{e317f}
\end{eqnarray}
\setcounter{equation}{\value{enumi}}
}
\noindent
The antisymmetric part of $F_{ij}(p)$ is, therefore, given by
{\addtocounter{equation}{0}
\setcounter{enumi}{\value{equation}}
\addtocounter{enumi}{1}
\setcounter{equation}{0}
\renewcommand{\theequation}{\thesection.\theenumi\alph{equation}}
\begin{eqnarray}
F_{12}^{(A)}(p)&=&\frac{1}{2}[S(x),P(x)]
                    =\frac{1}{2}if_{abc}T_c S^a(x)P^b(x),
\label{e318a} \\
F_{23}^{(A)}(p)&=&F_{31}^{(A)}(p)=0.
\label{e318b}
\end{eqnarray}
\setcounter{equation}{\value{enumi}}
}
\noindent
The third bosonic Lagrangian ${\cal L}_3$ in (\ref{e219c}) then becomes
\begin{eqnarray}
{\cal L}_3&=&\eta\,\displaystyle{\sum_p}
                {\rm tr}[F_{ij}^{(S)\dag}(x,p)F^{(S)ij}(x,p)]
               +\zeta\,\displaystyle{\sum_p}
                {\rm tr}[F_{ij}^{(A)\dag}(x,p)F^{(A)ij}(x,p)]\nonumber\\
	  &=&{\rm\ symmetric\ part\ }
               +\frac{1}{2}\zeta(f_{abc}S^a(x)P^b(x))^2.
\label{e319}
\end{eqnarray}

\noindent
The second bosonic Lagrangian ${\cal L}_2$ in (\ref{e219b}) is
\begin{eqnarray}
{\cal L}_2&=&\xi\,\displaystyle{\sum_{p,i}}
                {\rm tr}[(\nabla\!_\mu H(x,p,p+g_i))^{\dag}(\nabla^\mu H(x,p,p+g_i))\nonumber\\
            &=&\xi\,\displaystyle{\sum_p}
              {\rm tr}[\nabla\!_\mu H(x,p,p+g_1)\nabla^\mu H(x,p,p+g_1)
            +\nabla\!_\mu H(x,p,p+g_2)\nabla^\mu H(x,p,p+g_2)]\nonumber \\
            &=&2\xi[\nabla\!_\mu P^a(x)\nabla^\mu P_a(x)
            +\nabla\!_\mu S^a(x)\nabla^\mu S_a(x)].
\label{e320}
\end{eqnarray}

\noindent
The first bosonic Lagrangian ${\cal L}_1$ is the same as (\ref{e219a}).
The total bosonic Lagrangian ${\cal L}_1+{\cal L}_2+{\cal L}_3$ is, therefore,
identical with that in (\ref{e301}), only when $\eta=0$, $\zeta=-1$ and 
$\xi=\frac{1}{4}$.  

Thus we have obtained the $N=2$ super Yang-Mills theory from the viewpoint of 
the $M_4\times Z_2\times Z_2$ gauge theory.  In this interpretation the scalar 
field $S^a(x)$ and pseudoscalar field $P^a(x)$ have been regarded as gauge 
fields along two directions on $Z_2\times Z_2$.  The antisymmetric curvature 
for both scalar fields has been important in this construction.

\setcounter{equation}{0}
\addtocounter{section}{1}
\section*{\S\thesection.\ \ $N=4$ super Yang-Mills theory}

\indent

  The $N=4$ super Yang-Mills action\cite{Brink} is given by
\begin{eqnarray}
I_4&= \displaystyle{\int d^4x}\Bigl[&\!-\frac{1}{4}(F_{\mu\nu}^a)^2
	+\frac{1}{2}i\bar\chi^{aj}\gamma^\mu\nabla\!_\mu\chi_{aj}
	+\frac{1}{2}(\nabla\!_\mu A^{aI})^2
	+\frac{1}{2}(\nabla\!_\mu B^{aI})^2 \nonumber\\
   & &-if_{abc}\,\bar\chi^{aj}(\alpha_{jk}^I A^{bI}
	+i\gamma_5\beta_{jk}^I B^{bI})\chi^{ck} \nonumber\\
   & &-\frac{1}{4}\{(f_{abc}A^{bI}A^{cJ})^2
	+(f_{abc}B^{bI}B^{cJ})^2+2(f_{abc}A^{bI}B^{cJ})^2\}\Bigr],
\label{e401}
\end{eqnarray}

\noindent
where $\nabla\!_\mu$ is the gauge covariant derivative with $f_{abc}$ the 
structure constants of an arbitrary gauge group $G$.  
All fields belong to the adjoint representation of $G$, and there is 
a global $SU(4)$ internal symmetry with a central $SO(4)$ subgroup of scalar 
charges.  
The notation and classification of fields are given in Table.  
The $4\times 4$ matrices $\alpha_{jk}^I$ and $\beta_{jk}^I$ are coupling 
constant matrices of the $(1,0)$ or $(0,1)$ and $(\frac{1}{2},\frac{1}{2})$ 
irreps of $SO(4)$.  
They coincide with the $\eta$ and $\bar\eta$ matrices of instanton theory.

\vspace{.7cm}
{\bf Table.}\ \ Fields for $N=4$ supersymmetric Yang-Mills theory

\begin{tabular}{ccll} \hline
{\it Spin}&{\it Multiplicity}&{\it Fields}&{\it $SO(4)$ irrep.} \\ \hline
$1$           & $1$              & $A_\mu^a$  & $(0,0)$ \\
$\frac{1}{2}$ & $4$              & Majorana spinor $\chi^{aj}$, $j=1,\cdots,4$ & $(\frac{1}{2},\frac{1}{2})$ \\ 
$0^\pm$   & $3+3 $             & $A^{aI},B^{aI}$, $I=1,2,3$ & $(1,0)$ and $(0,1)$\\ \hline
\end{tabular} 
\vspace{.7cm}

  Our purpose is to derive the above action from the viewpoint of the 
$M_4\times Z_2\times Z_2$ gauge theory.  The procedure is quite the same as 
in the previous section.  Only difference is to introduce the coupling 
constants $\alpha_{jk}^I$ and $\beta_{jk}^I$ into the covariant differences 
(\ref{e307}) and (\ref{e308})
\begin{eqnarray}
\nabla\!_1\psi^j(x,p)=\psi^j(x,p)
                         -2\alpha_{jk}^I H^I(x,p,p+g_1)\psi^k(x,p+g_1),
\label{e402} \\
\nabla\!_2\psi^j(x,p)=\psi^j(x,p)
                         -2\beta_{jk}^I H^I(x,p,p+g_2)\psi^k(x,p+g_2),
\label{e403}
\end{eqnarray}

\noindent
where $(H^I(x,p,p+g_i))^a_{\,b}\psi^{bk}(x,p+g_i)$ is a parallel-transported vector of $\psi^{bk}(x,p+g_i)$ from $p+g_i$ to $p$.
In the following we set
\begin{eqnarray}
H^I(x,p,p+g_1)=B^I(x)=T_a B^{aI}(x),
\label{e404} \\
H^I(x,p,p+g_2)=A^I(x)=T_a A^{aI}(x),
\label{e405}
\end{eqnarray}

\noindent
where $T_a$ is subject to the algebra (\ref{e311}).  If we put fermionic fields
\begin{eqnarray}
\psi^{aj}(x,g_0)=\psi^{aj}(x,g_3)=\displaystyle{\displaystyle{\frac{L\chi^{aj}}
		{\sqrt{2}}}\choose0},
\label{e406} \\ \psi^{aj}(x,g_1)=\psi^{aj}(x,g_2)=\displaystyle{0\choose
		\displaystyle{\frac{R\chi^{aj}}{\sqrt{2}}}},
\label{e407}
\end{eqnarray}

\noindent
then the fermionic Lagrangian (\ref{e220}) becomes
\begin{equation}
{\cal L}_{\rm F}=\frac{1}{2}i\,\bar\chi^{aj}\gamma^\mu\nabla\!_\mu\chi_{aj}
                -\bar\chi^{aj}(\alpha^I_{\,jk}A^I
                +i\gamma_5\beta^I_{\,jk}B^I)_{ab}\chi^{bk},
\label{e408}
\end{equation}

\noindent
which is equivalent to that in (\ref{e401}) in the adjoint representation.

  Next we consider the bosonic Lagrangian.  
In the same way as in the $N=2$ case, we consider only antisymmetric 
curvatures for the discrete space.  
However, contrary to the $N=2$ case, there are three kinds of antisymmetric 
curvatures here.  
One of them corresponds to (\ref{e318a}), {\it i.e.},
\begin{equation}
F_{g_1g_2}(x,p)=[A^I(x),B^J(x)]=if_{abc}T_cA^{aI}(x)B^{bJ}(x).
\label{e409}
\end{equation}

\noindent
Geometrically, this is a difference between two routes of parallel 
transportations of $\psi(x,p+g_3)$ depicted in ${\rm Fig.C}_1$.  This curvature 
$F_{g_1g_2}(x,p)$ is the antisymmetric part of the triangle curvature 
corresponding to ${\rm Fig.C}_2$.

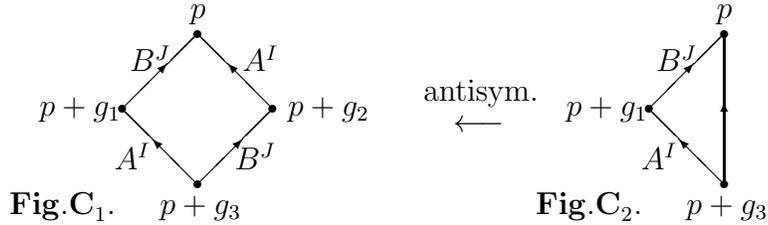
\setlength{\unitlength}{1mm}
\begin{picture}(165,40)(0,0)
 \put(40,10){\vector(-1,1){6}}\put(34,16){\line(-1,1){4}}
 \put(30,20){\vector(1,1){6}}\put(36,26){\line(1,1){4}}
 \put(40,10){\vector(1,1){6}}\put(46,16){\line(1,1){4}}
 \put(50,20){\vector(-1,1){6}}\put(44,26){\line(-1,1){4}}
 \put(40,10){\circle*{1}}
 \put(30,20){\circle*{1}}
 \put(50,20){\circle*{1}}
 \put(40,30){\circle*{1}}
 \put(39,32){$p$}
 \put(19,19){$p+g_1$}
 \put(52,19){$p+g_2$}
 \put(35,06){$p+g_3$}
 \put(29,12){$A^I$}
 \put(45,12){$B^J$}
 \put(31,25){$B^J$}
 \put(46,25){$A^I$}
 \put(15,06){${\rm\bf Fig.C}_1.$}
 \put(110,10){\vector(0,1){11}}\put(110,21){\line(0,1){9}}
 \put(110,10){\vector(-1,1){6}}\put(104,16){\line(-1,1){4}}
 \put(100,20){\vector(1,1){6}}\put(106,26){\line(1,1){4}}
 \put(110,10){\circle*{1}}
 \put(100,20){\circle*{1}}
 \put(110,30){\circle*{1}}
 \put(109,32){$p$}
 \put(89,19){$p+g_1$}
 \put(105,06){$p+g_3$}
 \put(99,12){$A^I$}
 \put(101,25){$B^J$}
 \put(85,06){${\rm\bf Fig.C}_2.$}
 \put(70,21){antisym.}
 \put(74,17){$\longleftarrow$}
\end{picture} 

\noindent
The other two are 
\begin{equation}
[A^I(x),A^J(x)] {\rm\ \  and\ \ } [B^I(x),B^J(x)].
\label{e410}
\end{equation}

\noindent
Both curvatures vanish in the $N=2$ case.  
However, in the $N=4$ case they don't vanish since they have $I,J$ components. 
Geometrically, $[A^I,A^J]$ corresponds to a difference between two routes 
of parallel transportations depicted in ${\rm Fig.D}_1$, {\it i.e.}, 
\begin{eqnarray}
(p+g_3\!\!&\to&\!\! A^J\to p+g_1\to 1\to p+g_2\to A^I\to p)\nonumber\\
&&-(p+g_3\to A^I\to p+g_1\to 1\to p+g_2\to A^J\to p),
\label{e411}
\end{eqnarray}

\noindent
where the mapping function $H(p+g_1,p+g_2)$ is unity from (\ref{e306}).  
The curvature $[A^I(x),A^J(x)]$ is the antisymmetric part of the triangle-like 
curvature corresponding to ${\rm Fig.D}_2$. The same is true for $[B^I(x),B^J(x)]$.

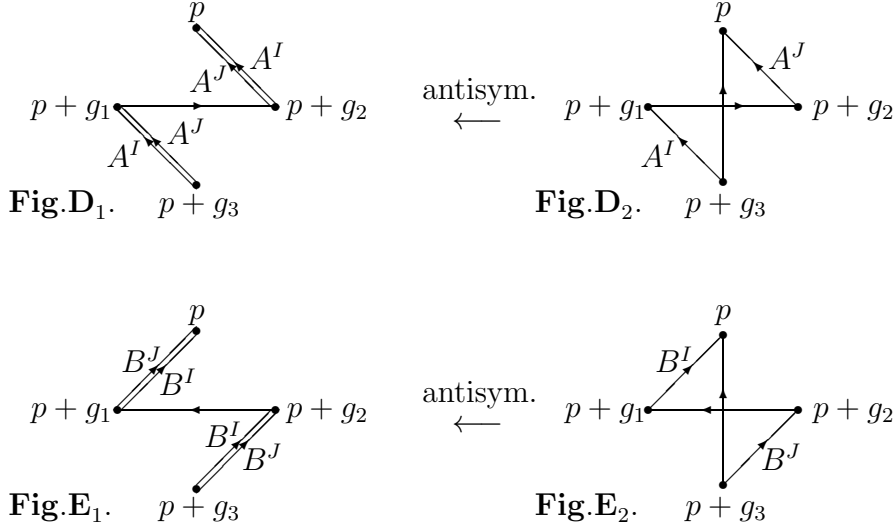
\setlength{\unitlength}{1mm}
\begin{picture}(165,40)(0,0)
 \put(40,10){\vector(-1,1){6}}\put(34,16){\line(-1,1){4}}
 \put(40,9){\vector(-1,1){7}}\put(33,16){\line(-1,1){4}}
 \put(30,20){\vector(1,0){11}}\put(41,20){\line(1,0){9}}
 \put(50,20){\vector(-1,1){6}}\put(44,26){\line(-1,1){4}}
 \put(51,20){\vector(-1,1){6}}\put(45,26){\line(-1,1){5}}

 \put(40,9.5){\circle*{1}}
 \put(29.5,20){\circle*{1}}
 \put(50.5,20){\circle*{1}}
 \put(40,30.5){\circle*{1}}
 \put(39,32){$p$}
 \put(18,19){$p+g_1$}
 \put(52,19){$p+g_2$}
 \put(35,06){$p+g_3$}
 \put(28,12){$A^I$}
 \put(36,15){$A^J$}
 \put(39,22){$A^J$}
 \put(47,25){$A^I$}
 \put(15,06){${\rm\bf Fig.D}_1.$}
  \put(110,10){\vector(0,1){13}}\put(110,23){\line(0,1){7}}
 \put(100,20){\vector(1,0){13}}\put(113,20){\line(1,0){7}}
 \put(110,10){\vector(-1,1){6}}\put(104,16){\line(-1,1){4}}
 \put(120,20){\vector(-1,1){6}}\put(114,26){\line(-1,1){4}}
 \put(110,10){\circle*{1}}
 \put(100,20){\circle*{1}}
 \put(120,20){\circle*{1}}
 \put(110,30){\circle*{1}}
 \put(109,32){$p$}
 \put(89,19){$p+g_1$}
 \put(122,19){$p+g_2$}
 \put(105,06){$p+g_3$}
 \put(99,12){$A^I$}
 \put(116,25){$A^J$}
 \put(85,06){${\rm\bf Fig.D}_2.$}
 \put(70,21){antisym.}
 \put(74,17){$\longleftarrow$}
\end{picture}  

\setlength{\unitlength}{1mm}
\begin{picture}(165,40)(0,0)
 \put(40,10){\vector(1,1){6}}\put(46,16){\line(1,1){4}}
 \put(40,9){\vector(1,1){7}}\put(47,16){\line(1,1){4}}
 \put(50,20){\vector(-1,0){11}}\put(39,20){\line(-1,0){9}}
 \put(30,20){\vector(1,1){6}}\put(36,26){\line(1,1){4}}
 \put(29,20){\vector(1,1){6}}\put(35,26){\line(1,1){5}}

 \put(40,9.5){\circle*{1}}
 \put(29.5,20){\circle*{1}}
 \put(50.5,20){\circle*{1}}
 \put(40,30.5){\circle*{1}}
 \put(39,32){$p$}
 \put(18,19){$p+g_1$}
 \put(52,19){$p+g_2$}
 \put(35,06){$p+g_3$}
 \put(30,25){$B^J$}
 \put(35,22){$B^I$}
 \put(46,12){$B^J$}
 \put(41,15){$B^I$}
 \put(15,06){${\rm\bf Fig.E}_1.$}
  \put(110,10){\vector(0,1){13}}\put(110,23){\line(0,1){7}}
 \put(120,20){\vector(-1,0){13}}\put(107,20){\line(-1,0){7}}
 \put(110,10){\vector(1,1){6}}\put(116,16){\line(1,1){4}}
 \put(100,20){\vector(1,1){6}}\put(106,26){\line(1,1){4}}
 \put(110,10){\circle*{1}}
 \put(100,20){\circle*{1}}
 \put(120,20){\circle*{1}}
 \put(110,30){\circle*{1}}
 \put(109,32){$p$}
 \put(89,19){$p+g_1$}
 \put(122,19){$p+g_2$}
 \put(105,06){$p+g_3$}
 \put(115,12){$B^J$}
 \put(101,25){$B^I$}
 \put(85,06){${\rm\bf Fig.E}_2.$}
 \put(70,21){antisym.}
 \put(74,17){$\longleftarrow$}
\end{picture}  

  The bosonic Lagrangian is, therefore, given by 
\begin{eqnarray}
{\cal L}_{\rm B}&=&-\frac{1}{4}(F_{\mu\nu}^a)^2
		+2\xi[(\nabla\!_\mu A^{aI})^2
                     +(\nabla\!_\mu B^{aI})^2]\nonumber\\
		& &-2\zeta{\rm tr}\{[A^I,A^J]^2+[B^I,B^J]^2+2[A^I,B^J]^2\}.
\label{e412}
\end{eqnarray}

\noindent
This is equivalent to the bosonic parts of (\ref{e401}) if 
$\xi=-\zeta=\frac{1}{4}$.  
Thus we have obtained the $N=4$ super Yang-Mills theory from the viewpoint 
of the $M_4\times Z_2\times Z_2$ gauge theory, when we used antisymmetric 
curvatures for scalar fields.  
Both scalar fields $A^{aI}(x)$ and $B^{aI}(x)$ have been regarded as gauge 
fields along two directions on $Z_2\times Z_2$. 

\setcounter{equation}{0}
\addtocounter{section}{1}
\section*{\S\thesection.\ \ Concluding remarks}

\noindent

  We have considered the $N=2$ and $4$ super Yang-Mills theories from the 
viewpoint of the $M_4\times Z_2\times Z_2$ gauge theory.  
The scalar fields $S^a(x)$, $P^a(x)$ in the $N=2$ case and $A^{aI}$, 
$B^{aI}$, $I=1,2,3$ in the $N=4$ case have been introduced as gauge fields 
along directions on $Z_2\times Z_2$ discrete space.  
The ``covariant derivatives" on the discrete space have given the Yukawa couplings between fermions and scalar fields.  

  The kinetic terms of these scalar fields and the Higgs potentials have been 
determined by curvatures which come from scalar fields.  
Here, the important things are that there are symmetric and antisymmetric 
curvatures for scalar fields.  
We have seen that only antisymmetric curvatures for scalar fields are related 
to the $N=2$ and $4$ super Yang-Mills theories.  

  There is no antisymmetric curvature for $Z_2$ or $Z_3$ discrete space.  
The $Z_2\times Z_2$ discrete space is the first space that includes such 
antisymmetric curvature.  
$Z_4$ is essentially the same as $Z_2\times Z_2$.  
This is the reason why we use $Z_2\times Z_2$.  
We have seen the geometrical meaning of symmetric and antisymmetric 
curvatures.  

  In the NCG formulation one can also define such symmetric and antisymmetric 
curvatures.  
To see this let us use Sitarz's one-form $\chi^i$\cite{Sitarz}.  
The two-form curvature is 
\begin{equation}
F=F_{ij}\chi^i\land\chi^j.
\label{e501}
\end{equation}

\noindent
where $F_{ij}$ corresponds to the triangle curvature (\ref{e211}) and 
$F_{ij}\ne\pm F_{ji}$.  
In general $\chi^i\land\chi^j\ne\pm\chi^j\land\chi^i$.  
The Lagrangian is given by the inner product 
\begin{eqnarray}
\langle F,F\rangle&=&{\rm tr}(F_{ij}^{\dag}F_{kl})
                     \langle\chi^i\land\chi^j,\chi^k\land\chi^l\rangle
                     \nonumber \\
                  &=&{\rm tr}(F_{ij}^{\dag}F_{kl})
                     [\langle \chi^i,\chi^k\rangle\langle\chi^j,\chi^l\rangle
		     +\langle\chi^i,\chi^l\rangle \langle\chi^j,\chi^k\rangle]
                     \nonumber \\
     &=&{\rm tr}(F_{ij}^{\dag}F_{kl})(a\,\delta^{ik}\delta^{jl}
				    +b\,\delta^{il}\delta^{jk})\nonumber \\
     &=&a\,{\rm tr}(F_{ij}^{\dag}F^{ij})
       +b\,{\rm tr}(F_{ij}^{\dag}F^{ji}),
\label{e502}
\end{eqnarray}

\noindent
where $a,b$ are generally arbitrary real constants.  
In the framework of NCG one cannot fix them definitely.  
So we have the symmetric curvature Lagrangian for $a=b$, the antisymmetric 
curvature Lagrangian for $a=-b$, and generally both mixed.  
The result (\ref{e502}) corresponds to our ${\cal L}_3$ in (\ref{e219c}).

  Finally we emphasize that a superspace formulation for $M_4 \times Z_2 \times
Z_2$ gauge theory may be one of the most important problems left unsolved. 
We expect that from this formulation the relationship between the 
antisymmetric curvature and the supersymmetry becomes clearer.  

\vspace{1.5cm}

\noindent
{\large \bf Acknowledgments}:
We thank G. Konisi and K. Shigemoto for useful discussions and 
invaluable comments.  
One of us (K.U.) is grateful to the special research funds at 
Tezukayama University. 


\end{document}